\documentclass[aps,prl,twocolumn,showpacs,preprintnumbers,superscriptaddress,groupedaddress]{revtex4}  
\usepackage{graphicx}  
\usepackage{dcolumn}   
\usepackage{bm}        
\usepackage{slashed}
\usepackage{amssymb} \usepackage{amsmath}   

\def\cN{\mathcal{N}}
\def\cM{\mathcal{M}}
\def\tr{\textrm{tr}}

\def\cMt{\overset{_0}{\mathcal{M}}}
\def\cMo{\overset{_1}{\mathcal{M}}}

\def\sq[#1,#2]{\left[#1\,#2\right]}
\def\an[#1,#2]{\left\langle#1\,#2\right\rangle}
\def\spab[#1,#2,#3]{\left\langle#1|#2|#3\right]}

\begin{document}
\preprint{IPhT-T14/108, IHES/P/14/33, ACFI-T14-21}
\title{Bending of Light in Quantum Gravity}
\author{N.~E.~J.~Bjerrum-Bohr}\email{bjbohr@nbi.dk}
\affiliation{Niels Bohr  International  Academy and  Discovery
Center,
The Niels Bohr Institute, Blegdamsvej 17,
DK-2100 Copenhagen \O, Denmark}
\author{John~F.~Donoghue}
\email{donoghue@physics.umass.edu}
\affiliation{Department of Physics-LGRT,
University of Massachusetts,
Amherst, MA 01003 USA}
\author{Barry~R.~Holstein} \email{holstein@physics.umass.edu}
\affiliation{Department of Physics-LGRT,
University of Massachusetts,
Amherst, Massachusetts 01003 USA}
\author{Ludovic~Plant\'e} \email{ludovic.plante@cea.fr}
\affiliation{CEA, DSM, Institut de Physique Th{\'e}orique, IPhT, CNRS, MPPU,
URA2306, Saclay, F-91191 Gif-sur-Yvette, France
}
\author{Pierre~Vanhove}\email{pierre.vanhove@cea.fr}
\affiliation{CEA, DSM, Institut de Physique Th{\'e}orique, IPhT, CNRS, MPPU,
URA2306, Saclay, F-91191 Gif-sur-Yvette, France
}
\affiliation{
Institut des Hautes {\'E}tudes Scientifiques,
Bures-sur-Yvette, F-91440, France }

\date{\today}

\begin{abstract}
We consider the scattering of lightlike matter in the presence of a heavy
scalar object (such as the Sun or a Schwarzschild black hole). By treating
general relativity as an effective field theory we directly compute
the nonanalytic components of the one-loop gravitational amplitude for the
scattering of massless scalars or photons from an external massive scalar
field. These results allow a semiclassical computation of the bending
angle for light rays grazing the Sun, including long-range $\hbar$ contributions.
We discuss implications of this computation, in particular the
violation of some classical formulations of the equivalence principle.
\end{abstract}

\maketitle

Since the discovery of quantum mechanics and general relativity in the previous
century it has been clear that these two theories have completely
different notions of reality at a fundamental level. While
deterministic physics  is a crucial ingredient in general relativity,
{\it i.e.}, particles follow field equations formulated as geodesic
equations, in quantum mechanics such a concept has no meaning since
one has to accept that space and momentum are mutually complementary
concepts.  The notion of a quantum field theory offers  a middle ground to some extent
by combining these concepts through field variables, but the
traditional formulation of such a theory suffers from
(nonrenormalizable) divergences in the ultraviolet regime.
Whatever the high-energy theory of gravity turns out to be,
it is intriguing that we can already answer a number of
important questions simply by employing an effective field
theory framework for general relativity, wherein the basic building block is
the Einstein-Hilbert Lagrangian.  In order to absorb ultraviolet
divergences we include in the action all possible invariants allowed by
the basic symmetries of the theory.  This infinite set of corrections is
usually seen as a signal of the loss of predictability and as a
dependence on the high-energy completion of the theory.
However, at one-loop order something surprising happens that was first
noticed by~\cite{Donoghue:1994dn} and was exploited in~\cite{BjerrumBohr:2002kt,
Bjerrum-Bohr:2013bxa}---the basic
Einstein-Hilbert term is sufficient to extract the long-range behavior of the theory.
This feature was used to extract the quantum corrections to the Newtonian
potential of a small mass attracted by a larger mass:
\vskip-0.55cm
\begin{equation}\nonumber
V(r) =- {G M m\over r} \bigg(1 + {3 G (M+m)\over { c^2 r}}+ {41 G\hbar\over {10 c^3 r^2}}\bigg)\,.
\end{equation}
\vskip-0.1cm\noindent
Here $M$ is a large (scalar) object, say the Sun, $m$ is a small test mass, $r$
is the distance between the two objects, and $G$, $c$ and $\hbar$, are Newton's constant, the
speed of light and the Planck constant respectively. Since these initial computations
there have appeared a number of papers computing
various potentials~\cite{Holstein:2008sx}, involving {\it e.g.},
fermionic and spin-1 matter.  It has been explicitly demonstrated
that the spin-independent components of one-loop general relativity theory display
universality both for the classical contribution as well as for the one-loop quantum
correction~\cite{Holstein:2008sx,Bjerrum-Bohr:2013bxa}.

In this Letter we will focus on a different problem, which has not yet been
discussed in the literature, namely computing the leading quantum correction
to the gravitational bending of light around the
Sun~\cite{BlackburnThesis}. Our goal is to show that this quantity is
readily calculable using modern field theory techniques. In doing
so we find that the quantum corrections do not respect classical
formulations of the equivalence principle. While the net effect is far
too small to be seen experimentally, this quantum violation
of the equivalence principle is an interesting phenomenon in its own right.

This Letter is organized as follows.  First we briefly review how to treat
general relativity as an effective field theory coupled to
photons and to lightlike (massless) scalar matter.  We work out amplitudes
for the gravitational scattering of the photons as well as of massless scalar matter as a reference.
As we will demonstrate, even at the quantum level of general relativity the
universality of the couplings to energy-momentum holds largely unchanged.
We show how our computation of the cross section can be used to
deduce a semiclassical deflection angle in which the post-Newtonian general
relativistic corrections are reproduced and new quantum mechanical corrections are generated.
Finally, we conclude and summarize our results.

We begin by considering the Einstein-Hilbert Lagrangian coupled to QED
and two neutral scalar fields
\vskip-0.52cm
\begin{eqnarray}\label{e:EHQED}
\hskip-0.9cm
  \mathcal S&=& \int d^4x \,
  \sqrt{-g}\,\bigg [\Big({2\over \kappa^2} \,\mathcal R -\frac14 (\nabla_\mu
    A_\nu-\nabla_\nu A_\mu)^2\Big) \cr
&+&\Big(-\frac12 (\partial_\mu \varphi)^2-\frac12
  ((\partial_\mu\phi)^2-M^2 \phi^2)\Big)+S_{\rm EF}\bigg ],
\end{eqnarray}
\vskip-0.3cm\noindent
where the covariant derivative is given by $\nabla_\mu A^\nu:=\partial_\mu
    A^\nu+\Gamma^\nu{}_{\mu\lambda} A^\lambda$ (we will be
      using the Feynman gauge) where  $\Gamma^\lambda{}_{\mu\nu}=1/2 \,
g^{\lambda\sigma}(\partial_\mu g_{\sigma\nu}+\partial_\nu
g_{\sigma\mu}-\partial_\sigma g_{\mu\nu}) $ are the Christoffel
symbols and $\kappa^2=32\pi G_N/c^4$.  The fields are denoted in the following way:
gravitons $h$, photons $\gamma$, massless scalars $\varphi$, and
massive scalars $\phi$. $S_{\rm EF}$ denotes the higher derivative contributions present
in an effective field theory.

In order to utilize this theory consistently, it is important to
consider it as an effective field theory\cite{Donoghue:1994dn}, by inclusion of a string of
higher-order operators in the action.  Divergences, being local, are absorbed into the coefficients
of these local higher-order operators. However, the long-range contributions correspond
to nonanalytic terms in momentum space or equivalently nonlocal behavior in
coordinate space.  These contributions are ultraviolet finite and follow uniquely from
the vertices of $S_{\text{EF}}$. For the purposes of evaluating
only the longest-range contributions, we need not display these higher order terms in the action.

The calculation is greatly simplified by two remarkable facts. One is that the on-shell gravitational
tree-level amplitudes can be obtained as the square of gauge theory
amplitudes~\cite{Kawai:1985xq,Bern:2002kj}. In our case
the gravitational Compton amplitudes will be reduced to the product of QED Compton amplitudes~\cite{Holstein:2006bh,Bjerrum-Bohr:2013bxa,BBHPV}.
The difficult calculations involving the triple graviton vertex can be avoided and are replaced by the much
simpler QED vertices.

%
\vskip-0.3cm
\begin{figure}[ht]
  \centering
  \includegraphics[width=6cm, height=2cm]{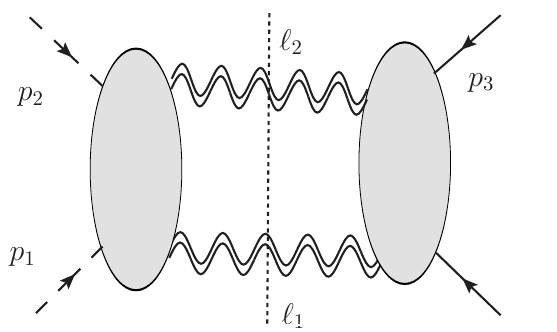}
  \vskip-0.35cm
  \caption{\small The two gravitons cut for the amplitude between a massless
    particle (dashed line) and  the massive scalar (solid line). The
    grey blobs are tree-level gravitational Compton amplitudes.}
  \label{fig:cut}
\end{figure}
\vskip-0.5cm
%

The other great simplification is to use on-shell unitarity techniques \cite{Bern:1994cg}, instead of Feynman diagrams. 
Calculating gravitational Feynman loops is a long and tedious process using the vertex rules of the gravitational Lagrangian.
Unitarity-based calculations construct the relevant amplitude from the discontinuity of the scattering process.  
The long-range nonanalytic terms in the one-loop amplitude can be readily calculated from these on-shell cuts using the
property of unitarity, as was directly demonstrated in
ref.~\cite{Bjerrum-Bohr:2013bxa}. Cutting the graviton internal lines
(see figure~\ref{fig:cut}), the integrand of the one-loop amplitude
factorizes in terms of a product of relatively simple tree amplitudes, given in our case by
the gravitational Compton amplitudes. 

The corresponding cut graviton exchange discontinuity amplitude (denoted disc.) takes the form
\vspace{-0.15cm}
\begin{eqnarray}
  \label{e:oneloop}\hskip-5cm &&
 i \cMo{}^{[\eta(p_1)\eta(p_2)]}_{[\phi(p_3)\phi(p_4)]} \Big|_{\rm disc}=\\ &&\hskip-0.5cm
  \int\!\!\!\frac{d^D\!\ell}{(2\pi)^4} \frac{\underset{\tiny h_1,h_2}{\sum} \!\! \cMt{}^{[h^{h_1}(\ell_1)h^{h_2}(-\ell_2)]}_{[\eta(p_1)\eta(p_2)]} \,
    \cMt{}^{[h^{h_1}(-\ell_1)h^{h_2}(\ell_2)]}_{[\phi(p_3)\phi(p_4)]}\,^\star}{4\ell_1^2
  \ell_2^2},\nonumber
\end{eqnarray}
\vspace{-0.0cm}\noindent
with the on-shell conditions $\ell_1^2=\ell_2^2=0$ for the cut
momenta of the internal  graviton lines.  For the photon case
$\eta=\gamma$ and for the massless scalar case $\eta=\varphi$.
Here $D=4-2\epsilon$ and the $\star$ denotes conjugation. Our notation
here is $\cMt$ for tree-level gravitational Compton amplitudes and
$\cMo$ for the one-loop amplitude of figure~\ref{fig:cut}. We follow the notation and momentum conventions of~\cite{Bjerrum-Bohr:2013bxa}
with all momenta defined as incoming and set $q:=p_1+p_2=-p_3-p_4=\ell_2-\ell_1$ with $t:=q^2=(p_1+p_2)^2$
and $p_3^2=p_4^2=M^2$.  In the ``all-incoming'' convention $t$ corresponds to the momentum transfer of a scattering process.

The relation  between the gravitational and
electrodynamic Compton processes is given by
\vspace{-0.2cm}
\begin{equation}\label{e:GravCompton}
 \hspace{-0.00cm} i \cMt{}^{[h(k_1)h(k_2)]}_{[\eta(p_1)\eta(p_2)]}= {\kappa^2\over 4e^2}\,{(p_1\cdot k_1)(p_1\cdot
    k_2)\over p_1\cdot p_2}\,  \cMt^{\rm QED}_{S=0}\, \cMt^{\rm QED}_{\eta}\!\!\!\!\,,
\end{equation}
\vskip-0.2cm\noindent
and is derived in detail in~\cite{BBHPV}.  Here $\cM^{\rm QED}_{\gamma}=\cM^{\rm QED}_{S=1}$ utilizes the Compton amplitude for the scattering of a photon from a massless charged spin-1 target
while $\cM^{\rm QED}_{\varphi}=\cM^{\rm QED}_{S=0}$ employs the Compton amplitude of a photon from a
massless charged spin-0 target.
These tree-level relations connect one-loop gravitational physics with one-loop electrodynamics in a nontrivial and interesting 
way~\cite{Bjerrum-Bohr:2013bxa}.

A final simplification is the use of the spinor-helicity formalism
(see~\cite{Mangano:1990by} for a review). While this notation is perhaps less familiar to
some, it drastically reduces the form of the amplitudes which we now display.
The only nonvanishing gravitational Compton helicity amplitudes involving photons $\gamma$ and gravitons $h$ are
\vskip-0.5cm
\begin{eqnarray}\label{e:ggGGhel}
\!\!\!\!\!\!\!\!\! i \cMt{}^{[h^+(k_1)h^-(k_2)]}_{[\gamma^+(p_1)\gamma^-(p_2)]}\!\!\!&=\!\!&{\kappa^2\over 4}{\sq[p_1,k_1]^2\an[p_2,k_2]^2\spab[k_2,p_1,k_1]^2\over
    (p_1\cdot p_2)(p_1\cdot k_1)(p_1\cdot k_2)},
\end{eqnarray}
\vskip-0.2cm\noindent
with $\cMt{}^{[h^+(k_1)h^-(k_2)]}_{[\gamma^-(p_1)\gamma^+(p_2)]}$
given by the above formula with $p_1$ and $p_2$ interchanged, and
amplitudes with opposite helicity configurations are obtained by
complex conjugation.
 For the tree-level massive scalar-graviton interaction amplitude we have
\vskip-0.55cm
\begin{eqnarray}
  \label{e:4pointGrav}
 i \cMt{}^{[h^+(k_1)h^+(k_2)]}_{[\phi(p_1)\phi(p_2)]}&=&{\kappa^2\over 4}\,
 {M^4 \sq[k_1,k_2]^4\over (k_1\cdot k_2)( k_1\cdot p_1)(k_1\cdot p_2)} \,, \cr
 i \cMt{}^{[h^-(k_1)h^+(k_2)]}_{[\phi(p_1)\phi(p_2)]}&=&{\kappa^2\over 4}\,
{ \spab[k_1,p_1,k_2]^2\spab[k_1,p_2,k_2]^2\over (k_1\cdot k_2)( k_1\cdot p_1)(k_1\cdot p_2)}\,.
\end{eqnarray}
\vskip-0.1cm\noindent
The tree-level amplitudes between the massless scalar $\varphi$ and
the graviton are obtained by setting $M=0$. Amplitudes with opposite helicity configurations are obtained by
complex conjugation.

The discontinuity integral of~\eqref{e:oneloop}  is given by the sum
of four box integrals with the same numerator factor
\vspace{-0.25cm}
\begin{equation}\begin{split}  \label{e:oneloops}
    i \cMo{}^{[\eta(p_1)\eta(p_2)]}_{[\phi(p_3)\phi(p_4)]} \Big|_{\rm disc}&=\\
  &\hskip-4.2cm-{\kappa^4\over 4 t^4}\sum_{h_1,h_2}\sum_{i=1}^2\sum_{j=3}^4
  \int \frac{d^D\ell}{(2\pi)^4} \,{\cN^{h_1 \, h_2}\over \ell_1^2\ell_2^2 (p_i\cdot\ell_1)(p_j\cdot\ell_1)}\,,
\end{split}\end{equation}
\vspace{-0.1cm}\noindent
where $h_1$ and $h_2$ denote the helicities $(+/-)$ of the exchanged gravitons in
the cut.
With this construction one captures all the $t$-channel massless
thresholds, which are the only terms of interest to us. The cut is
evaluated as in~\cite{Bjerrum-Bohr:2013bxa}, resulting in a very
simple answer due to the dramatic simplification of
the gravitational Compton tree-level amplitudes
in~\eqref{e:GravCompton}:
the singlet cut with $h_1=h_2=+$ or $h_1=h_2=-$ vanishes and
the nonsinglet cut is given by 
\vskip-0.6cm
\begin{equation}
  \cN^{+-}+\cN^{-+}= \Re\textrm{e}\, \Big[\big(\tr_-(\slashed \ell_1 \slashed p_1\slashed\ell_2\slashed
  p_3)\big)^4\Big]\,.
\end{equation}
\vskip-0.20cm\noindent
for the massless scalar-massive scalar amplitude and 
\vskip-0.40cm
\begin{equation}
\!\!\!\!\cN^{+-}\!\!+\!\cN^{-+}\!=
\!\Re\textrm{e}\bigg[\!{\big(\tr_-(\slashed \ell_2 \slashed p_2 \slashed\ell_1 \slashed p_3)\tr_+(\slashed\ell_2 \slashed p_3 \slashed\ell_1 \slashed p_1 \slashed p_3 \slashed p_2)\big)^2\over\langle p_1 | p_3 | p_2]^2}\bigg]\!\,,
\end{equation}
\vskip-0.20cm
\noindent
for the  photon-massive scalar amplitude where $\tr_\pm(\cdots):=\tr(
  {1\pm\gamma_5\over2}\cdots )$.
Performing standard tensor integral reductions~\cite{Bern:1992em} into scalar boxes,
scalar bubbles and
scalar triangle integrals, the amplitude is decomposed
in terms of integral functions with a massless $t$-channel cut
\vskip-0.50cm
\begin{eqnarray}
&&\hskip-.8cm  -{1\over 4\kappa^4} \cMo{}^{[\eta(p_1)\eta(p_2)]}_{[\phi(p_3)\phi(p_4)]} \Big|_{\rm disc}
 = bo^\eta(t,u)   \, I_4(t,u)+
  bo^\eta(t,s)\, I_4(t,s)\cr
&+& t_{12}^\eta(t) \, I_3(p_1,p_2,0)+ t_{34}^\eta(t)\, I_3(p_3,p_4,M^2)\cr
&+& bu^\eta(t) \, I_2(t,0)\,.
\end{eqnarray}
\vskip-0.10cm\noindent
Here $I_4(t,u)$ and $I_4(t,s)$ are the scalar box integrals given in
\S 4.4.6 of~\cite{Ellis:2007qk}, $I_3(t)$ is
the massless triangle integral with vanishing internal masses, and
$I_3(t,m)$ is the finite massive triangle integral and $I_2(t)$ is the
massless scalar bubble integral both given in Eq.~(III.17)
of~\cite{Bjerrum-Bohr:2013bxa}. [In the massless ($M\to0$) limit this computation reproduces the
graviton cut given by Dunbar and Norridge; see eq.~(4.10)~\cite{Dunbar:1995ed}.]

The integral reduction yields massive bubbles as well as tadpoles and
analytic pieces that do not possess a massless $t$-channel cut.  Such
pieces are not completely determined from the cut and are not of
interest to our analysis since they do not contribute to the long-range 
interactions at low-energy.

Computation of the cut discontinuity can be accomplished using traditional methods and is greatly simplified by the use of on-shell identities.  We will elsewhere present the details of these computations and here quote only the leading (nonanalytic) results required to perform the analysis of the
cross section and the semiclassical bending angle.

In the leading low-energy ($\omega\ll M$) limit, where $\omega$ is the
frequency of the photon,  the  total amplitude sum of the tree-level
and one-loop contributions
$i \cM =  {i\over \hbar}\, \cMt+   i \cMo$
 takes the very striking form
 \vspace{-0.3cm}
  \begin{eqnarray}
 &&    \cM^{[\eta(p_1)\eta(p_2)]}_{[\phi(p_3)\phi(p_4)]}
 =-{\cN^{\eta}\over\hbar}\, \Big[
\kappa^2\,  {(2M\omega)^{2}\over 4 t}\cr
&+&\hbar {\kappa^4\over 4} \, \Big(
4(M\omega)^4 (I_4(t,u)+I_4(t,s))+ 3(M\omega)^2 t I_3(t)\nonumber\\[.5ex]
&-&15 (M^2\omega)^2
I_3(t,M)+ bu^{\eta} (M\omega)^2 I_2(t)
 \Big)\Big]\,,
 \end{eqnarray}
where $\cN^{\varphi}=1$ for the massless scalar, while for
the photon, we have $\cN^{\gamma}=(2M\omega)^{2} / (2\langle
p_1|p_3|p_2]^2)$ for the $(+-)$ photon helicity contribution and its complex
conjugate for the $(-+)$ photon helicity contribution.
The photon amplitude vanishes for the polarization configurations $(++)$ and
$(--)$ is a direct consequence of the properties of the tree amplitudes in
eq.~\eqref{e:ggGGhel} .
Notice that $|\cN^{\gamma}|^2\to1$ in the low-energy
limit and that this prefactor does not affect the cross section.
The coefficients of the bubble contributions are $bu^{\varphi}= 3/40$ and $bu^{\gamma}=-161/120$.

It is a striking example of the universality of the gravitational
couplings that all coefficients--except those for the
bubble--are identical (for the leading contribution) for the scalar and photon
scattering.

From this result we can compute the leading contribution to the
amplitude (expanding all integrals in terms of leading-order
contributions as done in~\cite{BjerrumBohr:2002kt,Bjerrum-Bohr:2013bxa}):\vspace{-0.1cm}
\
\begin{eqnarray} \label{e:oneloopsInt}
&&     \cM^{[\eta(p_1)\eta(p_2)]}_{[\phi(p_3)\phi(p_4)]}
 \simeq{\cN^{\eta}\over \hbar}\,  (M\omega)^2\\
&&\times\nonumber
 \Big[{\kappa^2\over t}-\kappa^4 {15\over
  512}{ M\over \sqrt{-t}} \\ &&\nonumber
-\hbar \kappa^4
 {15\over
  512\pi^2}\,\log\left(-t\over M^2\right)
+\hbar\kappa^4\,{ bu^{\eta} \over(8\pi)^2} \, \log\left(-t\over \mu^2\right)
\\
\nonumber &&-\hbar\kappa^4 {3\over128\pi^2}\, \log^2\left(-t\over \mu^2 \right)
-\kappa^4 \,  {M\omega\over 8\pi} {i\over t}\log\left(-t\over
  M^2\right)
\Big]\,.
\end{eqnarray}
\vskip-0.1cm\noindent
where $\mu^2$ is an arbitrary mass scale parameter used in dimensional regularization.
The two terms in the second line correspond, respectively, to the
leading Newtonian contribution and the first post-Newtonian
correction~\cite{Donoghue:1994dn,BjerrumBohr:2002kt,Bjerrum-Bohr:2013bxa}. The
next three logarithmic terms represent quantum gravity
corrections. The first term on the third line corresponds to
the quantum correction to the metric evaluated
in~\cite{BjerrumBohr:2002ks}.  The second term on the third line
arises from the one-loop ultraviolet
divergence of the amplitude and is the only contribution depending
on the spin of the massless field.  On the fourth line the first term,
involves a new form not found in the previous analysis.
Finally, the last term, arising from the discontinuity
of the box integral, contributes to the phase
of the amplitude and is not directly observable. For this reason it
will not be considered further.

It is very interesting that in the low-energy limit the one-loop amplitudes
for the massless scalar and for the photon involve the same coefficients
except for the $bu^{\eta} \,\log(-t/\mu^2)$ contribution from the massless bubble. This means that
these massless particles feel the same gravitational interaction from the
massive object except for this quantum contribution.  Since the matter content and properties are different for the
scalar and photonic theories,  obtaining a universal result for the
bubble coefficient should not be expected.
[The arguments in~\cite{Holstein:2008sx,Bjerrum-Bohr:2013bxa} imply
that the amplitude for a massless spin-$\frac12$ scattering on the
Sun will differ as well only by the bubble contribution.]

Note that, because of the vanishing of the photon scattering amplitudes
for the helicity configurations $(++)$ and $(--)$, the amplitude is the same
in the plane of scattering or along its orthogonal component, which explicitly rules out the
possibility of birefringent effects.

We do not know of a fully quantum treatment of the bending of light which is capable of describing the one-loop amplitude.
However, in order to try to understand
the impact of the above corrections, we can proceed by defining,
in the small momentum transfer limit $t\simeq -{\vec q}^{\,2}$, a semiclassical potential for a massless scalar and photon
interacting with a massive scalar object by use of the Born approximation
\vskip-0.6cm
\begin{eqnarray}\nonumber
V_\eta(r)\!\!&=&\!\!{\hbar\over 4M\omega}\!\! \int\!\!
\cM^{[\eta(p_1)\eta(p_2)]}_{[\phi(p_3)\phi(p_4)]}(\vec q)\, e^{i \vec
  q\cdot r}\! {d^3q\over (2\pi)^3}\cr
&\hspace{-0.8cm}\simeq& -{2G M \omega\over r}-{15\over 4}{(GM)^2 \omega\over r^2}\cr
&-& {8 bu^\eta-15\over 4\pi}\,{G^2 M \omega\hbar\over r^3}
- {12 G^2 M \omega\hbar\over \pi}\,{\log{r\over r_o}\over r^3}\,.
\end{eqnarray}
\vskip-0.1cm\noindent
where $r_0$ is an infrared scale.

Using na\"ively the semiclassical formula for angular deflection given 
in chap.~21~\cite{bohm}and in~\cite{Donoghue:1986ya} and the form of the above
potential we find for the bending angle of a photon and for a massless scalar
\vspace{-0.15cm}
\begin{eqnarray}\label{e:theta}
&&\theta_\eta \simeq- {b\over \omega}\int_{-\infty}^{+\infty} {V_\eta'(b \sqrt{1+u^2})\over
  \sqrt{1+u^2}} du\\
\nonumber &\simeq& {4G M\over b}+{15\over 4} {G^2 M^2 \pi \over
  b^2}
+{8 bu^\eta+9-48 \log {b\over 2r_o}\over \pi}\,{G^2 \hbar M\over  b^3}\,.
\end{eqnarray}
\vskip-0.15cm\noindent
The first two terms give the correct classical values, including the first
post-Newtonian correction, expressed in terms of the gauge-invariant
impact parameter $b$ (see, for
instance~\cite{Willangle}).  The last term is a quantum gravity effect
of the order $G^2\hbar M/b^3=\ell_P^2
r_S/(2b^3)$ which involves the product of the Planck length and the
Schwarzschild radius of the massive object divided by the cube of the impact
parameter.

The quantum effect depends on the spin of massless particle scattering
on the massive target.
Of course, this dependence does not necessarily violate the equivalence principle,
in that the logarithmic quantum corrections correspond to nonlocal effects in coordinate
space. Because of the long-distance propagation of massless photons and gravitons in
loops, such quantum effects are not localized, and the
difference can be interpreted as a tidal correction in that the massless particle can no longer
be described as a point particle. There is no requirement from the equivalence principle that
such nonlocal effects be independent of the spin of the massless particle. Nevertheless, we see that particles no longer travel
on geodesics and that different particles bend differently. This is certainly
in contrast to classical applications of the equivalence principle.

{Let us compare} the bending angle
of a photon with that of a massless scalar by the Sun. The only difference
given by the above treatment will
be given by the bubble effect
\vskip-0.75cm
\begin{equation}
  \theta_\gamma-\theta_\varphi =  {8 (bu^\gamma-bu^\varphi)\over \pi} \,
  {G^2\hbar M\over b^3}  
\end{equation}
\vskip-0.15cm\noindent
and is far too small to be seen experimentally~\cite{Will:2005va}.
However, it is interesting that quantum effects do predict such a difference,
modifying one of the key features of classical general relativity.
Moreover, this is another demonstration that effective field techniques can make
well-defined predictions within quantum gravity.
We have focused on the bending of light in the vicinity of a massive object.
One can envision other situations wherein
the effective field theoretic framework might be very useful to analyze and
understand effects in quantum gravity.  We find such a prospect indeed to be very exciting!

\smallskip
\begin{acknowledgments}
We thank Zvi Bern, Simon Caron-Huot, Poul Henrik
Damgaard, Thibault Damour, Paolo Di Vecchia,  C\'edric Deffayet, Basem El-Menoufi, Niels Obers,
Ugo Moschella and Ciaran Williams for useful discussions. We thank
Thibault Damour for pointing out  incorrect signs in a previous version
of this paper.
 N.E.J.B. thanks the Institute des Hautes \' Etudes Scientifiques
for hospitality during the preparation of this work.  The research of P.V. was supported by the Agence Nationale de la 
Recherche grant 12 BS05 003 01, and the Centre National 
de la Recherche Scientifique Grants Projets Internationaux
de Collaboration Scientifique number 6076 and 6430.
The research of J.F.D. was supported by the US National Science Foundation grant PHY-1205986.
\end{acknowledgments}

\end{document}